\documentclass[11pt,a4paper]{article}

\usepackage[T1]{fontenc}
\usepackage[utf8]{inputenc}
\usepackage{lmodern}
\usepackage{microtype}
\usepackage[a4paper,margin=26mm,headheight=15pt]{geometry}

\usepackage{amsmath,amssymb,amsthm,mathtools,bm,mathrsfs}
\usepackage{booktabs,array,enumitem}
\usepackage{xcolor}

\usepackage[colorlinks=true,
            linkcolor=blue!45!black,
            citecolor=blue!45!black,
            urlcolor=blue!45!black,
            pdfdisplaydoctitle=true]{hyperref}
\numberwithin{equation}{section}
\newtheorem{theorem}{Theorem}[section]
\newtheorem{proposition}[theorem]{Proposition}
\newtheorem{lemma}[theorem]{Lemma}
\newtheorem{corollary}[theorem]{Corollary}
\theoremstyle{definition}
\newtheorem{definition}[theorem]{Definition}
\newtheorem{example}[theorem]{Example}
\theoremstyle{remark}
\newtheorem{remark}[theorem]{Remark}
\newtheorem{openproblem}[theorem]{Open Problem}

\usepackage{cleveref}
\crefname{theorem}{Theorem}{Theorems}
\crefname{proposition}{Proposition}{Propositions}
\crefname{lemma}{Lemma}{Lemmas}
\crefname{corollary}{Corollary}{Corollaries}
\crefname{definition}{Definition}{Definitions}
\crefname{example}{Example}{Examples}
\crefname{remark}{Remark}{Remarks}
\crefname{openproblem}{Open Problem}{Open Problems}
\newcommand{\cH}{\mathcal H}
\newcommand{\cK}{\mathcal K}
\newcommand{\cA}{\mathcal A}
\newcommand{\cB}{\mathcal B}
\newcommand{\cC}{\mathcal C}
\newcommand{\cM}{\mathcal M}
\newcommand{\cN}{\mathcal N}
\newcommand{\cR}{\mathcal R}
\newcommand{\BH}{\mathcal B(\mathcal H)}
\newcommand{\Even}{\mathrm{Even}}
\newcommand{\id}{\mathbf 1}
\newcommand{\vN}{\overline\otimes}

\newcommand{\sot}{\xrightarrow{\mathrm{s}}}

\newcommand{\Rea}{\operatorname{Re}}
\newcommand{\mer}{\beta_{\mathrm M}}
\newcommand{\III}{\mathrm{III}}
\newcommand{\II}{\mathrm{II}}

\def\<{\langle}

\def\C{{\mathbb{C}}}\def\H{{\mathcal{H}}}
\def\i{{\rm{i}}}

\def\>{\rangle}

\newcolumntype{L}[1]{>{\raggedright\arraybackslash}p{#1}}
\setlist[enumerate]{itemsep=3pt,topsep=5pt}
\allowdisplaybreaks[1]
\hypersetup{%
  pdftitle={Maximal Violation of the Mermin Inequality for Commuting von Neumann Algebras},
  pdfauthor={Kan He, Jinchuan Hou, Shuyuan Yang, Bingke Zheng},
  pdfsubject={Bell nonlocality in algebraic quantum field theory},
  pdfkeywords={Bell nonlocality, Mermin inequality, von Neumann algebras, type III factors, strong stability}}

\title{Maximal Violation of Mermin Inequality for Commuting
       von Neumann Algebras and Multipartite Vacuum Entanglement in QFT}
\author{%
  Kan He\textsuperscript{1,2},\ 
  Jinchuan Hou\textsuperscript{1},\ 
  Shuyuan Yang\textsuperscript{3,4,5}, and
  Bingke Zheng\textsuperscript{1}%
  \thanks{The authors are listed in alphabetical order of their surnames.}\\[10pt]
  {\small
   \parbox{152mm}{\centering
     \textsuperscript{1} College of Mathematics, Taiyuan University of Technology,
       Taiyuan 030024, China\\[1pt]
     \textsuperscript{2}School of Mathematics and Statistics, Shaanxi Normal University,
       Xi'an 710119, China\\[1pt]
     \textsuperscript{3}School of Mathematics, North University of China,
       Taiyuan 030051, China\\[1pt]
     \textsuperscript{4}School of Cyberspace Science and Technology,
       Beijing Institute of Technology, Beijing 100081, China\\[1pt]
     \textsuperscript{5}Center for Quantum Technologies, National University of Singapore,
       Singapore 117543, Singapore\\[5pt]
     E-mail: \texttt{hekanquantum@163.com},\ 
     \texttt{jinchuanhou@aliyun.com},\ 
     \texttt{yangshuyuan2000@163.com},\
     \texttt{zhengbingke02@163.com}}}}
\date{}

\begin{document}
\maketitle
\pagestyle{headings}

\begin{abstract}
Summers and Werner proved that Bell inequalities are violated maximally and
generically
in the vacuum of a relativistic quantum field theory in [S. J. Summers and R. Werner, Commun. Math. Phys. 110 (1987), 247–259]. Their criterion concerns the bipartite case, that is, there are two algebras of quantum observables: one algebra and its
commutant. In this case, the maximal CHSH value can be 
approached arbitrarily closely in every normal
state exactly when the algebra is strongly stable. In this paper, we study the
corresponding question for three or more mutually commuting von Neumann
algebras. We focus on the multipartite Mermin inequalities instead of CHSH ones. For the tripartite Mermin functional we show that maximal
violation in every normal state is equivalent to the existence of a
common sequence of asymptotic GHZ stabilizers. Strong stability of each
local factor is a necessary condition for maximal violation, and a joint embedding of an infinite GHZ
tensor product is sufficient. Counterexamples show that strong stability and type $\III_1$ of the individual algebras, even together with tripartite duality, do not suffice to guarantee maximal violation in every normal state.
The results  are also extended to $n\ge3$
parties.

\par\smallskip
\end{abstract}

\section{Introduction}
\label{sec:intro}

Bell's theorem~\cite{Bell} shows that  local hidden-variable models cannot reproduce all correlations 
predicted by quantum theory, and the
CHSH inequality~\cite{CHSH} is its canonical two-party form. In the
normalization used below every local model satisfies $\beta_2\le1$,
while suitable measurements on a maximally entangled pair give
$\beta_2=\sqrt2$, the Tsirelson bound~\cite{Cirelson}. In the
nonrelativistic setting the observables available to each party
generate a full matrix algebra on a tensor factor of the total Hilbert
space, and the theory of Bell nonlocality is developed inside this
tensor product algebra (TPA) model (see~\cite{BCPSW} for a survey).

Relativistic quantum field theory calls for a different description.
In the algebraic framework of Haag and Kastler~\cite{HK} every
spacetime region $O$ is assigned a von Neumann algebra
$\mathfrak A(O)$ of observables localized in $O$, and the algebras of
spacelike separated regions commute. For large classes of models and
regions these local algebras are of type $\III$, and very often of type
$\III_1$. Araki~\cite{Araki} proved the type $\III$ property for the
free field, and~\cite{Sorce} is a recent account of the classification.
Type $\III$ factors admit no nonzero normal semifinite trace and have no minimal projections 
or normal pure states. The interpretation of measurements in this setting is
itself delicate~\cite{FewsterVerch}, and we refer to~\cite{KR,Takesaki}
for the general theory of operator algebras. The model in which the
observables of the parties are mutually commuting subalgebras of a
single $\BH$ is strictly more general than the TPA model, as the
negative solution of the Tsirelson problem~\cite{JNPPSW,MIPRE} shows,
and Bell inequalities must therefore be studied in the commuting-operator framework as well.

Summers and Werner~\cite{SW85,SW87a,SW87b,SW88,SW} discovered that this
re-examination leads not to a weakening but to a sharpening of Bell
nonlocality. If $\cM$ is a type $\III_1$ factor with a cyclic and
separating vector, then every vector state violates the CHSH inequality
maximally. Moreover the value $\sqrt2$ is attained in \emph{every}
normal state exactly when $\cM$ is strongly stable, that is,
$\cM\cong\cM\vN\cR$ (\cite{SW}). Thus, for the algebra–commutant configurations in QFT to which these results apply, the vacuum exhibits maximal CHSH violation. When the local algebra is strongly stable, maximal violation holds in every normal state, showing that this property is determined by the algebra–commutant configuration. Two ingredients of that proof recur below: 1) a
standard-form hypothesis, namely a common cyclic and separating vector,
which makes modular theory and the homogeneity of the state space of a
type $\III_1$ factor~\cite{CS} available, and 2) a structural condition on
the algebra, namely the absorption of the hyperfinite $\II_1$ factor
$\cR$~\cite{MvN}, a notion rooted in the classification of
Connes~\cite{Connes73,Connes76} and in McDuff's work on central
sequences~\cite{McDuff}. Quantitatively, Landau~\cite{Landau} showed
that the maximal CHSH value can be approached only if both local
algebras contain a copy of $M_2(\mathbb{C})$.

The multiparty case is qualitatively different. The GHZ
argument~\cite{GHZ} and Mermin's inequality~\cite{Mermin} exhibit a
three-party correlation that no local hidden-variable model can
reproduce: the operator $T$ of \eqref{eq:mermin3} satisfies
$\omega(T)\le1$ in every local model, while $T|GHZ\>=2|GHZ\>$ with
$|GHZ\>=(|000\rangle+|111\rangle)/\sqrt2$, an eigenvalue relation for
the four stabilizers $R_{j,m}$ of Definition~\ref{def:stabilizers}.
The $n$-party generalizations, usually called MABK after Mermin,
Ardehali and Belinskii--Klyshko~\cite{Mermin,Ardehali,BK}, give
violations that grow exponentially with $n$, whereas other
multipartite inequalities detect different notions of nonlocality. 
Svetlichny's inequality~\cite{Svetlichny} separates genuine three-party
nonlocality from biseparable mixtures, its $n$-party
refinements~\cite{SS} play the same role for $n\ge4$, and the
high-dimensional CGLMP family~\cite{CGLMP} tests the dimensional
content of the correlations. In this paper we work with the real-part
Mermin polynomial, whose normalization is fixed in \eqref{eq:constants}. This
 coincides with the MABK normalization for odd $n$ and is strictly
weaker for even $n$.

The closest precedents concern the Mermin inequality itself in QFT. De
Fabritiis, Roditi and Sorella~\cite{DRS2023} construct, from smeared
Dirac fields, unitary operators that create GHZ-type states out of the
Minkowski vacuum and show that the Mermin inequality is maximally
violated on those states, and the same circle of ideas is developed with
Weyl operators, modular theory and vacuum projectors in
\cite{DGGRS2024,GRSV2025}. These papers exhibit a maximally violating
state in specific models, whereas the question here is which
configurations of algebras give maximal violation in every normal state.
On the algebraic side, Halvorson and Clifton~\cite{HC2000} show that
Bell correlated states are generic for two commuting algebras of
infinite type, complementing the bipartite criterion recalled above,
while~\cite{PWPWJ2008} shows that tripartite Bell violations, unlike
bipartite ones, can be unbounded. In the tensor product model the
maximal violation of MABK self-tests the GHZ
state~\cite{Kaniewski2017,Panwar2023}, the finite-dimensional analogue
of the asymptotic stabilizer condition used below;
information-theoretic uses of the MABK value are made
in~\cite{Grasselli2021,Zhao2026}, and the optimal MABK violations are
determined in~\cite{Wu2026}. To our knowledge, the state-independent
version of the question has not been considered for three or more
algebras.

All of the results quoted above  about multipartite Bell violations in QFT concern a \emph{fixed} state. They
exhibit a state, or a family of states, on which a given inequality is
violated beyond its local bound. They do not answer the question that
the bipartite criterion of Summers and Werner answers, namely whether
maximal violation is a property of the configuration of algebras alone,
independent of the state.

The present paper addresses this question for the Mermin inequalities. We first build the multipartite Mermin inequalities for commuting von Neumann algebras. Furthermore, we focus on the question of maximal violation in every normal state of Mermin inequalities. The
first observation is that the bipartite correspondence
\[
 \text{maximal violation in every normal state}
 \quad\Longleftrightarrow\quad
 \text{strong stability of the algebra}
\]
cannot hold verbatim. We give a new necessary and sufficient condition for all-normal-state maximal violation, and explore the relation between strong stability and maximal violation.

The paper is organized as follows. Section~\ref{sec:review} recalls
the bipartite criterion of Summers and Werner and fixes the notation.
Section~\ref{sec:tripartite} 
builds the Mermin inequality on three commuting algebras. Section 4 treats the tripartite maximal violation in every normal state of Mermin inequalities. Section~\ref{sec:npartite}
extends the analysis to $n\ge3$ parties.
Section~\ref{sec:interpretation} discusses some further questions.

\section{Preliminaries}
\label{sec:review}

Throughout, Hilbert spaces are separable, the von Neumann algebras under
consideration have the same identity, and observables are self-adjoint
contractions. Let $\cR$ denote the hyperfinite type $\II_1$ factor. An
algebra $\cM$ is called \emph{strongly stable} if
$\cM\cong\cM\vN\cR$. We use the following results from the original
paper as established theorems; we do not reprove the underlying results
of Connes.

\begin{theorem}[Summers and Werner, \cite{SW}]\label{thm:SW}
Let the von Neumann algebra $\cM\subset\BH$ have a cyclic and separating vector, and define
\[
 \beta_2(\omega;\cM,\cM')
 =\sup_{\{A_x, B_y\}}\frac12|\omega\bigl(A_0(B_0+B_1)+A_1(B_0-B_1)\bigr)|,
\]
where $\omega$ is a state on $\BH$,  $A_x\in\cM$ and $B_y\in\cM'$ are self-adjoint contractions.
\begin{enumerate}[label=(\roman*)]
\item If \(\mathcal M\) is a type \(\mathrm{III}_1\) factor, then
\(\beta_2(\omega_\xi;\mathcal M,\mathcal M')=\sqrt{2}
\)
for every unit vector \(\xi\in\mathcal H\), where
\(
\omega_\xi(X)=\langle\xi,X\xi\rangle,\, X\in\mathcal B(\mathcal H).
\)
\item The algebra $\cM$ is strongly stable if and only if
$\beta_2(\omega;\cM,\cM')=\sqrt2$ for every normal state $\omega$
on $\BH$.
\end{enumerate}
\end{theorem}

{\it Connection with bipartite vacuum entanglement.} The link is provided by the Reeh--Schlieder theorem and by the type
$\mathrm{III}_1$ structure of local algebras.  If $O$ is a region with
nontrivial causal complement, the vacuum vector $\Omega$ is cyclic and
separating for $\mathcal A(O)$, so the vacuum is not a product state
across the cut $(\mathcal A(O),\mathcal A(O)')$: it is entangled between
$O$ and its complement.  Summers and Werner showed that this entanglement
is maximal in the operational sense~\cite{SW87b,SW}.  For a local algebra
of type $\mathrm{III}_1$---the case for free fields and for large classes
of regions in interacting theories---the vacuum violates the CHSH
inequality to the Tsirelson bound ($\sqrt2$ in the normalisation
in which local models give $1$), and when the local algebra is strongly
stable, as the hyperfinite type $\mathrm{III}_1$ factors of free fields and
of large classes of interacting models are, the same maximal violation
holds in every normal state, not merely in the vacuum.  The clause
``every normal state'' is precisely what connects the result to vacuum
entanglement. It shows that the nonlocality is a property of the algebra
inclusion rather than of the
particular vector, so the vacuum is not a fine-tuned entangled
state but one instance of a structural, state-independent entanglement
encoded in the type $\mathrm{III}_1$ nature of the local algebras. Under the standard-form and strong-stability hypotheses, maximal Bell violation is a structural property of the algebra--commutant configuration and holds in every normal state, including the vacuum state.

\section{Tripartite Mermin correlations}
\label{sec:tripartite}

We first establish the tripartite Mermin inequality on the model of commuting von Neumann algebras. Assume three von Neumann algebras 
$\cM_A,\cM_B, \cM_C\subset\BH$ satisfy
\[
 [\cM_A,\cM_B]=[\cM_A,\cM_C]=[\cM_B,\cM_C]=0.
\]
Let $A_x\in\cM_A$, $B_y\in\cM_B$, and $C_z\in\cM_C$, with
$x,y,z\in\{0,1\}$, be self-adjoint contractions. 
Define the normalized Mermin operator
\begin{equation}\label{eq:mermin3}
 T=\frac12\left(
 A_0B_0C_0-A_0B_1C_1-A_1B_0C_1-A_1B_1C_0\right)
\end{equation}
and
\[
 \mer(\omega,\cM_A,\cM_B,\cM_C)=\sup_{\{A_x, B_y, C_z\}}|\omega(T(A_x, B_y, C_z))|.
\]
In the tensor-product algebra model $\cM_A\otimes \cM_B \otimes \cM_C$, if $\beta_M$ is larger than 1, then the state $\omega$ is multipartite entangled. In the commuting algebra model, we will verify the multipartite entanglement by violation of the Mermin functional $\mer$, too.

Next we determine the upper and lower bounds of the normalized Mermin functional. 
We write a new notation for the Mermin functional as $\mer(\omega,\cM_A,\cM_B,\cM_C, k)=\mer(\omega, k)$, where $k$ denotes the number of abelian algebras. We have the following conclusion. 

\begin{theorem}\label{thm:tripartite}
For a 
state $\omega$, $1\leq\mer(\omega,\cM_A,\cM_B,\cM_C, k)\leq b(\omega, k)$, and 
\begin{center}

\begin{equation}\label{eq:tripartite}
 \sup_{\omega}b(\omega, k)=
 \begin{cases}
 2, & k=0,\\
 \sqrt2, & k=1,\\
 1, & k=2,3.
 \end{cases}
\end{equation}
\end{center}
\end{theorem}
Moreover, the bounds 1, \(\sqrt2\) and 2 are attainable.

\begin{proof}

\emph{Case $k=0$.}  Every product $A_xB_yC_z$ has norm at most 1. Hence  the triangle inequality gives 
$\|T\|\le\frac12\cdot4=2$, and therefore $\mer(\omega, 0)\le 2$. 

\emph{Case $k=1$.}  Assume without loss of generality that $\cM_{A}$ is
abelian. By character decomposition of von Neumann algebras \cite{KR}, there are real numbers $a_0(\chi), a_1(\chi)$ for some character
$\chi$ such that 
\[
 T=\frac12\Bigl(a_0(\chi)\bigl(B_0C_0-B_1C_1\bigr)
              -a_1(\chi)\bigl(B_0C_1+B_1C_0\bigr)\Bigr)
\]
on the fibre over $\chi$, with $|a_0(\chi)|,|a_1(\chi)|\le1$.   Hence a standard discussion shows that 
\[
 \|T\|\le\sqrt{a_0(\chi)^2+a_1(\chi)^2}\le\sqrt2 .
\]
Integrating over $\chi$ gives $|\omega(T)|\le\sqrt2$. So $\mer(\omega, 1)\leq \sqrt2$ in the case. 

\emph{Case $k=2$.}  Assume $\cM_{A}$ and $\cM_{B}$ are abelian.  By the character decomposition theory again, for  joint character $\chi$,
\[
 T=\frac12\Bigl(\bigl(a_0b_0-a_1b_1\bigr)C_0
              -\bigl(a_0b_1+a_1b_0\bigr)C_1\Bigr),
\]
where $a_x=\chi(A_x)$, $b_y=\chi(B_y)$.  The functional
\[
 \frac12 (|a_0b_0-a_1b_1|+|a_0b_1+a_1b_0|)
\]
is convex for every variable. The expression is convex in each variable separately, so its maximum
over $[-1,1]^4$ is attained at a vertex. At every vertex, one of
$|a_0b_0-a_1b_1|$ and $|a_0b_1+a_1b_0|$ equals $2$ and the other
equals $0$. Thus
\[
\frac{1}{2}\bigl(
|a_0b_0-a_1b_1|+|a_0b_1+a_1b_0|
\bigr)\le 1,
\]
and hence $\|T\|\le 1$. We get $\mer(\omega, 2)\le1$.

\emph{Case $k=3$.}  All three algebras are abelian, so $T$ is a scalar
for every joint character; the classical bound $|T|\le1$ and $\mer(\omega, 3)\le1$.

Finally, we show the bounds are tight. 
\emph{Case $k=0$.} Denote by $X,Y,Z$ the three Pauli matrices.  On $(\mathbb C^2)^{\otimes3}$, with $A_{j,0}=X$,
$A_{j,1}=Y$ and $\omega$ the GHZ state, one has
$T|GHZ\rangle=2|GHZ\rangle$,  hence $\mer(\omega, 0)=2$.

\emph{Case $k=1$.}  Take On \(\mathcal H=\mathbb C^2\otimes\mathbb C^2\), set
\(
\mathcal M_A=\mathbb CI_{\mathcal H},
\,
\mathcal M_B=M_2(\mathbb C)\otimes I_2,
\,
\mathcal M_C=I_2\otimes M_2(\mathbb C).
\)
Choose
\(
A_0=A_1=I_{\mathcal H},
\)
\(
B_0=\frac{X-Z}{\sqrt2}\otimes I_2,
\,
B_1=-\frac{X+Z}{\sqrt2}\otimes I_2,
\)
\(
C_0=I_2\otimes X,\, C_1=I_2\otimes Z.
\)
For
\(
\Phi=\frac{|00\rangle+|11\rangle}{\sqrt2},\,
\omega(D)=\langle\Phi,D\Phi\rangle,
\)
a direct calculation gives \(T\Phi=\sqrt2\Phi\). Therefore \(\beta_M(\omega, 1)=\sqrt2\).

\emph{Case $k=2,3$.}  Take $A_0=A_1=B_0=B_1=C_0=I$ and $C_1=-I$.
Then
\[
 T=\frac12\bigl(I+I+I-I)=I,
\]
so $\omega(T)=1$ for every state, hence $\mer(\omega, 2)=\mer(\omega, 3)=1$.
This completes the proof. \end{proof}

\begin{table}[htbp]
\centering
\small
\renewcommand{\arraystretch}{1.3}
\begin{tabular}{@{}clll@{}}
\toprule
$k$ & which algebras are abelian & sharp bound  & attained by \\
\midrule
$0$ & none & $2$ & GHZ, $X/Y$ settings \\
$1$ & one (any) & $\sqrt2$ & Bell pair on the other two parties \\
$2$ & any two & $1$ & $A_0=A_1=B_0=B_1=C_0=\id$, $C_1=-\id$ \\
$3$ & all three & $1$ & a deterministic classical assignment \\
\bottomrule
\end{tabular}
\caption{The tripartite Mermin correlation.  The quantum bound $2$ is
reached only when all three algebras are nonabelian, and one abelian algebra
already reduces it to $\sqrt2$, and two abelian algebras force the
classical bound $1$.}
\label{tab:tripartite}
\end{table}

\section{All-normal-state maximal violation of Mermin inequality} \label{maxviolation}

In this section, we want to explore the conditions for maximal violation of Mermin inequalities on commuting von Neumann algebras, which will generalize partially the Summers and Werner's results in \cite{SW}. The following lemma is helpful throughout the paper, which shows that the supremum of Mermin functional keeps invariant on unitary self-adjoint contractions. 

\begin{lemma}\label{lem:symmetries}
Let $\cM_A,\cM_B,\cM_C\subset \cB(\cH)$ be pairwise commuting von
Neumann algebras with a common identity. For each party put
\[
 M_A=\{A\in\cM_A:A^*=A\},\qquad K_A=\{A\in M_A:\|A\|\le1\},
\]
and define $M_B,K_B,M_C,K_C$ analogously. For a normal state $\omega$ on
$
\mathcal M_{ABC}
=\mathcal M_A\vee\mathcal M_B\vee\mathcal M_C
$ and the Mermin operator
\[
 T(A_0,A_1,B_0,B_1,C_0,C_1)=\tfrac12\bigl(A_0B_0C_0-A_0B_1C_1-A_1B_0C_1-A_1B_1C_0\bigr),
\]
the following statement holds
\[
 \sup_{\substack{ A_i\in K_A \\ B_j\in K_B \\ C_k\in K_C}}|\omega\bigl(T(A_0,A_1,B_0,B_1,C_0,C_1)\bigr)|
 =
 \sup_{\substack{A_i\in\operatorname{ext}K_A \\ B_j\in\operatorname{ext}K_B \\ C_k\in\operatorname{ext}K_C}}|\omega\bigl(T(A_0,A_1,B_0,B_1,C_0,C_1)\bigr)|.
\]
Moreover $\operatorname{ext}K_A=\{A\in K_A:A^2=\id\}$, i.e., the extreme
points of $K_A$ are precisely the self-adjoint unitaries of
$\cM_\cA$, and similarly for $K_B$ and $K_C$. Consequently, in computing
$\beta_M(\omega)$ the self-adjoint contractions may be restricted to
self-adjoint unitaries.
\end{lemma}

\begin{proof}
Let $M_A=\{A\in\cM_A:A^*=A\}$ and $K_A=\{A\in\cM_A:A^*=A,\|A\|\le1\}$,  
and define $M_B,M_C,K_B,K_C$ analogously. On bounded subsets of $M_A$, the $\sigma$-weak topology coincides with the weak-operator topology. We work with the $\sigma$-weak topology here. 

By the Alaoglu-Bourbaki theorem,
the unit ball of $\cB(\cH)$ is $\sigma$-weakly compact, i.e.,$\{T\in \cB(\cH):\|T\|\le1\}$ is $\sigma$-weakly compact.

Note that $K_A$ is a subset of the unit ball: the self-adjointness condition $A=A^{*}$ is $\sigma$-weakly closed (the adjoint map is continuous), and the condition $\|A\|\le 1$ is also closed. Hence $K_A$ is a closed subset of a compact set, therefore $\sigma$-weakly compact  and so $K_A$ is an ultraweakly closed subset of the unit ball of $\cB(\cH)$, and is therefore compact. It is clearly non-empty and convex. The same holds for $K_B$ and $K_C$.

Next we claim that $\text{ext}K_A=\{A\in K_A:A^2=I\}.$ First, we show that self-adjoint unitaries are extreme. Assume that $U\in K_A$ with $U^2=I$. Since $U$ is self-adjoint with $U^2=I$, its spectrum lies in $\{\pm1\}$. Let $P_{\pm}$ be the spectral projections onto the $\pm1$ eigenspaces, so $U=P_+-P_-$. Suppose $U=\lambda S+(1-\lambda)T$ with $\lambda\in(0,1)$ and $S,T\in K_A$. For a unit vector $\xi\in P_+H,$ we have $U\xi=\xi$, hence
$$1=\<\xi,U\xi\>=\lambda\<\xi,S\xi\>+(1-\lambda)\<\xi,T\xi\>.$$
Since $\<\xi,S\xi\>\le 1$ and $\<\xi,T\xi\>\le1$, a weighted average equal to 1 forces both to equal 1. This gives $S\xi=\xi$, and similarly $T\xi=\xi$. Hence $S|_{P_+H}=T|_{P_+H}=I$. The same argument on $P_-H$ gives $S=T=-I$. Therefore $S=T=U$, so $U$ is extreme.

Next, we show the non-unitary self-adjoint contractions are not extreme. Let $A\in K_A$ with $A^2\neq I$. Since $A$ is self-adjoint with $\|A\|\le1$, its spectrum lies in $[-1,1]$. If the spectrum were contained in $\{\pm1\}$, we would have $A^2=I$, a contradiction. Hence there exists a spectral point $t\in(-1,1)$. Choose \(\delta>0\) such that
\(
[t-\delta,t+\delta]\subset(-1,1),
\)
and set
\(
P=E_A((t-\delta,t+\delta))\neq0.
\)
For sufficiently small \(\varepsilon>0\), both \(A+\varepsilon P\) and \(A-\varepsilon P\) belong to \(K_A\). and 
$$A=\frac{1}{2}[(A+\varepsilon P)+(A-\varepsilon P)]$$
is a non-trivial convex decomposition, so $A$ is not extreme.


Denote $\hat\beta_{\rm M}=\sup_{\substack{A_i\in K_A\\B_j\in K_B\\C_k\in K_C}}|\omega(T(A_0,A_1,B_0,B_1,C_0,C_1))|.$
Fix $A_1,B_0,B_1,C_0,C_1$ and define $f(A_0):=|\omega(T(A_0,A_1,B_0,B_1,C_0,C_1))|,\ A_0\in K_A.$
With the other five observables fixed, the map $A_0\mapsto\omega(T)$ is affine and ultraweakly continuous. Its absolute value is therefore a continuous convex function on $K_A$. By Bauer's maximum principle,
$$\sup_{A_0\in K_A}f(A_0)=\sup_{A_0\in \text{ext}K_A}f(A_0).$$
Substituting the six variables $A_0,A_1,B_0,B_1,C_0,C_1$ successively, we obtain
$$\hat\beta_{\rm M}=\sup_{\substack{A_i\in\text{ext}K_A\\B_{j}\in\text{ext}K_B\\C_k\in\text{ext}K_C}}|\omega(T(A_0,A_1,B_0,B_1,C_0,C_1))|.$$

\end{proof}

To check all-normal-state violation, we introduce the following definition. 
\begin{definition}\label{def:stabilizers} Let \(\mathcal M_A,\mathcal M_B,\mathcal M_C\subset\mathcal B(\mathcal H)\) be pairwise commuting von Neumann algebras, and set
\(
\mathcal M_{ABC}
=(\mathcal M_A\vee\mathcal M_B\vee\mathcal M_C)''.
\)
For each \(m\in\mathbb N\), let
\(
A_{x,m}\in\mathcal M_A,\,
B_{y,m}\in\mathcal M_B,\,
C_{z,m}\in\mathcal M_C,
\, x,y,z\in{0,1},
\)
be self-adjoint unitaries. Define the following four self-adjoint unitaries on \(\mathcal M_{ABC}\):
\begin{align}
R_{1,m}&=A_{0,m}B_{0,m}C_{0,m},&
 R_{2,m}&=-A_{0,m}B_{1,m}C_{1,m},\\
 R_{3,m}&=-A_{1,m}B_{0,m}C_{1,m},&
 R_{4,m}&=-A_{1,m}B_{1,m}C_{0,m}.
\end{align}
We say that this sequence of local observable tuples satisfies the asymptotic GHZ stabilizer relations if
\(
R_{j,m}\xrightarrow{\mathrm{s}}I,\, j=1,2,3,4.
\)
The associated Mermin operators are
\(
T_m=\frac12\sum_{j=1}^{4}R_{j,m}.
\)
\end{definition}

The following theorem  establishes the 
equivalence of all-normal-state maximality and a common GHZ sequence.

\begin{theorem}\label{thm:Mequiv}
For a triple of pairwise commuting von Neumann algebras $\cM_A, \cM_B, \cM_C$ on a separable Hilbert space $\cH$, the following
conditions are equivalent:
\begin{enumerate}[label=(\roman*)]
\item Every normal state $\omega$ on $\cM_{\rm ABC}=(\cM_A \vee \cM_B\vee \cM_C)''$ satisfies $\mer(\omega)=2$.
\item There is a sequence of Mermin operators constructed from local
self-adjoint unitaries such that they converge to $2 I$ in the strong operator topology.
\item There is a sequence of local self-adjoint unitaries satisfying
the asymptotic GHZ stabilizer relations.
\end{enumerate}
 Replacing strong convergence in (ii) by ultraweak convergence gives
an equivalent condition.
\end{theorem}
\begin{proof}
(i)$\Rightarrow$(ii): 
Since $\cH$ is separable, choose an orthonormal basis $\{e_k\}_{k\ge1}$ and positive numbers $\{p_k\}$ with 
$p_k>0,\ \sum_{k\ge1}p_k=1$. Define 
$$\rho(D)=\sum_{k\ge1}p_k\<e_k,De_k\>,\ \ D\in B(\cH).$$
This is a normal state on $\cB(\cH)$. Now we verify the faithfulness of $\rho$.
If $\rho(D^{\dag}D)=0$, then for every $k$,
$$0\le p_k\<e_k,D^{\dag}D e_k\>\le\rho(D^{\dag}D)=0.$$
Since $p_k>0$, we get $\|De_k\|^2=\<e_k,D^{\dag}De_k\>=0,$ so $De_k=0$ for all $k$. Since $\{e_k\}$ is a basis, $D=0$. Hence $\rho$ is faithful.

The restriction of $\rho$ to $\mathcal M_{ABC}$ is a normal state.
Thus condition (i) gives $\beta_M(\rho)=2$, where we retain the
notation $\rho$ for its restriction. By Lemma \ref{lem:symmetries},
\[
\sup_T |\rho(T)|=2,
\]
where the supremum is taken over all Mermin operators constructed
from local self-adjoint unitaries.

Replacing both observables $A_0$ and $A_1$ by their negatives
replaces $T$ by $-T$ and preserves admissibility. Since $T$ is
self-adjoint, $\rho(T)$ is real. Consequently,
\[
\sup_T \rho(T)=\sup_T |\rho(T)|=2.
\]
We may therefore choose a sequence $(T_m)$ of such Mermin operators
satisfying
\[
2-\frac{1}{m}<\rho(T_m)\le 2.
\]
In particular, $\rho(T_m)\to 2$.

Let $E_m=2I-T_m$, then $T_m^{\dag}=T_m$, so $E_m^{\dag}=E_m$. And $\|T_m\|\le\frac{1}{2}(1+1+1+1)=2.$
Since $T_m$ is self-adjoint, $-2I\le T_m\le 2I$. Hence $0\le E_m=2I-T_m\le 4I$, and $\|E_m\|\le4.$

Since $\rho(T_m)\to 2$, we have
\[
\rho(E_m)=2-\rho(T_m)\to 0.
\]
As $E_m\ge 0$, every term in the expansion
\[
\rho(E_m)=\sum_{j\ge 1}p_j\langle e_j,E_me_j\rangle
\]
is nonnegative. Hence, for each fixed $k$,
\[
0\le p_k\langle e_k,E_me_k\rangle
\le \rho(E_m).
\]
Since $p_k>0$, it follows that
\[
0\le \langle e_k,E_me_k\rangle
\le \frac{\rho(E_m)}{p_k}\longrightarrow 0.
\]

By $0\le E_m\le4I$, the spectrum of $E_m$ lies in $[0,4],$ hence 
$E_m^2\le4E_m.$ Therefore, for each fixed $k$, 
$$\|E_me_k\|^2=\<e_k,E_m^2e_k\>\le4\<e_k,E_me_k\>\rightarrow0$$
implying $\|E_m e_k\|\rightarrow0\ \ (\text{for  each fixed}\ k)$.

Let $\xi=\sum_kc_ke_k\in \cH$ be arbitrary, with $\sum_k|c_k|^2=\|\xi\|^2<\infty.$ Let $\varepsilon>0$.
Since $\sum_k|c_k|^2<\infty,$ there exists $K$ such that 
$$(\sum_{k>K}|c_k|^2)^{\frac{1}{2}}<\frac{\varepsilon}{8}.$$
Set
$$\xi_{\leq K}=\sum_{k\leq K}c_ke_k,\ \ \ \xi_{>K}=\xi-\xi_{\le K}=\sum_{k>K}c_ke_k,$$
so that $\|\xi_{>K}\|<\frac{\varepsilon}{8}.$ And $\xi_{\leq K}$ is a finite sum. By $\|E_me_k\|\rightarrow0$ for each $k\le K$, one obtains $$\|E_m\xi_{\le K}\|\le\sum_{k\le K}|c_k|\|E_me_k\|\rightarrow0.$$
Hence there exists $N$ such that for $m>N,\ \|E_m\xi_{\le K}\|<\frac{\varepsilon}{2}.$ 
Since $\|E_m\|\le4,$ 
$$\|E_m\xi_{>K}\|\le \|E_m\|\|\xi_{>K}\|<4\times\frac{\varepsilon}{8}=\frac{\varepsilon}{2}.$$
So for $m>N$,
$$\|E_m\xi\|\le\|E_m\xi_{\le K}\|+\|E_m\xi_{>K}\|<\frac{\varepsilon}{2}+\frac{\varepsilon}{2}=\varepsilon.$$
Hence $E_m\xi\rightarrow0.$ Since $\xi$ is arbitrary,
$$E_m\xrightarrow{s}0$$
implying $T_m=2I-E_m\xrightarrow{s}2I$. Moreover, $T_m$ is built from local self-adjoint unitaries. This proves $(ii)$.

(ii)$\Rightarrow$(iii)
By definition, $T_m=\frac{1}{2}(R_{1,m}+R_{2,m}+R_{3,m}+R_{4,m}).$
Hence $$2I-T_m=2I-\frac{1}{2}\sum_{j=1}^{4}R_{j,m}=\frac{1}{2}\sum_{j=1}^4(1-R_{j,m}).$$ And it is easy to see that $R_{j,m}^{\dag}=R_{j,m},\ R_{j,m}^2=I.$
The spectrum of a self-adjoint unitary lies in
$\{-1,1\}$, so
$R_{j,m}\le I,\ I-R_{j,m}\ge0.$

By the condition $(ii)$, $T_m\xrightarrow{s}2I.$
By linearity of strong convergence, $2I-T_m\xrightarrow{s}0.$ For any fixed vector $\xi\in H$, strong convergence gives
$$\<\xi,(2I-T_m)\xi\>\rightarrow 0.$$
Then $\frac{1}{2}\sum_{j=1}^4\<\xi,(1-R_{j,m})\xi\>=\<\xi,(2I-T_m)\xi\>\rightarrow0,$ which implies $$\<\xi,(I-R_{j,m})\xi\>\rightarrow0,\ \ j=1,2,3,4$$ 
by using  $\<\xi,(I-R_{j,m})\xi\>\ge0$.

Since $R_{j,m}$ is self-adjoint and unitary, $R_{j,m}^2=I$, hence
$$(I-R_{j,m})^2=I-2R_{j,m}+R_{j,m}^2=2(1-R_{j,m}).$$
Therefore, for each fixed $j$ and $\xi$,
$$\|(I-R_{j,m})\xi\|^2=2\<\xi,(I-R_{j,m})\xi\>\rightarrow0.$$
Hence, $\|(I-R_{j,m})\xi\|\rightarrow0$ for every fixed $\xi$, i.e.,
$R_{j,m}\xi\rightarrow\xi$ for every $\xi$, which implies $R_{j,m}\xrightarrow{s}I,\ j=1,2,3,4.$

(iii)$\Rightarrow$(ii) follows by summation.

If all four $R_{j,m}$ converge strongly to $I$, then 
$$T_m=\frac{1}{2}(R_{1,m}+R_{2,m}+R_{3,m}+R_{4,m})\xrightarrow{s}2I.$$

(ii)$\Rightarrow$(i) follows from normality and the upper bound.

If $T_m\xrightarrow{s}2I$, then $\{T_m\}$ is uniformly bounded, hence $T_m\xrightarrow{\text{ultraweak}}2I$. Normal states are ultraweakly continuous, so for every normal state $\omega$, $\omega(T_m)\rightarrow2.$ Therefore, $2\ge \beta_M(\omega)
\ge \lim_{m\to\infty}|\omega(T_m)|
=2$. Hence $\mer(\omega)=2$.

Finally, we verify the assertion concerning ultraweak convergence.
Since $\|T_m\|\le 2$, strong convergence of $T_m$ to $2I$
implies ultraweak convergence.

Conversely, suppose that $T_m\to 2I$ ultraweakly, and set
$E_m:=2I-T_m$. As each $T_m$ is self-adjoint and
$\|T_m\|\le 2$, we have
\[
0\le E_m\le 4I,
\qquad
E_m^2\le 4E_m.
\]
For every $\xi\in H$, ultraweak convergence gives
$\langle \xi,E_m\xi\rangle\to 0$. Therefore,
\[
\|E_m\xi\|^2
=\langle \xi,E_m^2\xi\rangle
\le 4\langle \xi,E_m\xi\rangle
\longrightarrow 0.
\]
Thus $E_m\to 0$ strongly, and hence $T_m\to 2I$ strongly.
This proves the final assertion.

\end{proof}

Theorem \ref{thm:Mequiv} characterizes configurations for which the Mermin functional has supremum \(2\) in every normal state. We next examine the structural consequences of this property. The following theorem shows that, when the local algebras are factors, each of them must be strongly stable.

\begin{theorem}\label{thm:necessary3}
Suppose that $\cM_A,\cM_B,\cM_C$ are factors and that the triple \((\mathcal M_A,\mathcal M_B,\mathcal M_C)\) satisfies the equivalent conditions of Theorem~\ref{thm:Mequiv}. Then each factor is strongly
stable:
\[
 \cM_A\cong\cM_A\vN\cR,\quad
 \cM_B\cong\cM_B\vN\cR,\quad
 \cM_C\cong\cM_C\vN\cR.
\]
\end{theorem}
\begin{proof}
We prove the assertion for $\cM_\cA$, the other two parties can be similarly dealt with.

By condition (iii) of Theorem \ref{thm:Mequiv}, there exist six sequences of local self-adjoint unitaries
$$A_{i,m}\in \cM_A,\ B_{j,m}\in \cM_B,\ C_{k,m}\in\cM_C,\ \ i,j,k\in\{0,1\},$$
such that $R_{j,m}\xrightarrow{s}I,\ j=1,2,3,4,$ where $R_{1,m}=A_{0,m}B_{0,m}C_{0,m},\ R_{3,m}=-A_{1,m}B_{0,m}C_{1,m}$. Since $A_{0,m}^2=I, \ A_{1,m}^2=I,$ $\|A_{0,m}\|=1,$  and $\|A_{1,m}\|=1$, the left multiplication by $A_{0,m}$ and $A_{1,m}$, respectively, yields

\begin{equation}\label{eq:mirror}
 A_{0,m}-B_{0,m}C_{0,m}\sot0,
 \qquad A_{1,m}+B_{0,m}C_{1,m}\sot0.
\end{equation}

Since $B_{0,m}C_{0,m}\in \cM_{A}'$ and $-B_{0,m}C_{1,m}\in\cM_{A}'$, set $$D_{0,m}:=B_{0,m}C_{0,m}\in\cM_{A}',\ \ D_{1,m}:=-B_{0,m}C_{1,m}\in\cM_{A}'.$$
Thus $A_{0,m}-D_{0,m}\xrightarrow{s}0,\ A_{1,m}-D_{1,m}\xrightarrow{s}0$. Note that $A_{i,m}$ and $D_{i,m}$ are Hermitian, hence the convergence also holds in the strong $*$ topology.

We now use  the following general observation. 
Let \((A_m)\subseteq\mathcal M_A\) and \((D_m)\subseteq\mathcal M_A'\) be uniformly bounded sequences of self-adjoint operators such that
\(
E_m:=A_m-D_m\xrightarrow{\mathrm{s}}0.
\)
For every vector functional
 $\varphi_\xi(y)=\<\xi,y\xi\>$, and every $A\in \cM_A,$ since $D_m\in \cM_A'$, $$A_mA-AA_m=E_mA-AE_m.$$ Hence
\[
 \sup_{\substack{A\in\cM_A\\\|A\|\leq1}}
 |\varphi_\xi(A_mA-AA_m)|
 \leq2\|\xi\|\,\|E_m\xi\|\longrightarrow0.
\]
The linear span of vector functionals is norm dense in the predual $(\cM_\cA)_*$,
so $\|A_m\varphi-\varphi A_m\|\to0$ for every $\varphi\in(\cM_\cA)_*$.
Thus \eqref{eq:mirror} shows that $(A_{0,m})$ and $(A_{1,m})$ are
\emph{centralizing sequences}, i.e.,$\|A_{i,m}\varphi-\varphi A_{i,m}\|\rightarrow0,\ \forall \varphi\in (\cM_\cA)_*,\ i=0,1$.

On the other hand, since \(R_{j,m}\xrightarrow{\mathrm{s}}I\) for each \(j\) and the sequences are uniformly bounded,
\(
R_{1,m}R_{3,m}\xrightarrow{\mathrm{s}}I,\,
R_{4,m}R_{2,m}\xrightarrow{\mathrm{s}}I.
\)
Consequently,
\[
R_{1,m}R_{3,m}-R_{4,m}R_{2,m}
=-C_{0,m}C_{1,m}\{A_{0,m},A_{1,m}\}
\xrightarrow{\mathrm{s}}0.
\]
As $C_{0,m}C_{1,m}$ is unitary,
left multiplication by the inverse unitary preserves vector norms,
hence $\{A_{0,m},A_{1,m}\}\sot0$. The anticommutator is
self-adjoint, so this is again strong $*$ convergence.

Fix a free ultrafilter $\nu$. Define $$\mathscr C_\nu(\cM_A)=\{(a_m):\sup_m\|a_m\|<\infty,\ \lim_{m\rightarrow\nu}\|a_m\varphi-\varphi a_m\|=0,\forall \varphi\in (\cM_A)_*\}.$$
$$\mathscr I_\nu(\cM_A)=\{(a_m)\in\mathscr C_\nu(\cM_A):\lim_{m\rightarrow\nu}\|a_m\xi\|=0,\ \forall \xi\in H\},$$
and  $(\cM_A)_\nu:=\mathscr C_\nu(\cM_A)/\mathscr I_\nu(\cM_A),$ 
where $\mathscr C_\nu(\cM_A)$ is the algebra of bounded sequences
satisfying
$\lim_{m\to\nu}\|a_m\varphi-\varphi a_m\|=0$ for every
$\varphi\in(\cM_A)_*$. Its ideal $\mathscr I_\nu(\cM_A)$ consists of
those sequences converging to zero along $\nu$ in the $\sigma$-strong
$*$ topology.

Note that $(A_{0,m})_m$ and $(A_{1,m})_m$ lie in $\mathscr C_\nu(\cM_\cA)$: by \eqref{eq:mirror} each of them differs from the bounded sequence $D_{i,m}\in\cM_\cA'$ by a sequence tending strongly to $0$. Since $A_{i,m}^2=I$ and $A_{i,m}^*=A_{i,m}$ for every $m$, their images $a_i=[A_{i,m}]$ in the quotient satisfy
$a_i^2=I,\ a_i^*=a_i$.
Since $(\{A_{0,m},A_{1,m}\})\in \mathscr I_\nu(\cM_\cA),$ we have $\{a_0,a_1\}=0.$
Then $(a_0a_1)^2=-a_0^2a_1^2=-I$. If $a_0a_1=0$, then $(a_0a_1)^2=0$, a contradiction. Hence $a_0a_1\neq0$, and $a_0a_1\neq a_1a_0$. Therefore $(\cM_\cA)_\nu$ is nonabelian.
Connes' criterion~\cite{Connes} gives $\cM_\cA\cong\cM_\cA\vN\cR$. Separability ensures
the separable-predual hypothesis required by that criterion.\end{proof}

\begin{remark} The converse of Theorem \ref{thm:necessary3} 
 does not hold. Indeed, there exist three pairwise commuting, strongly stable type \(\mathrm{III}_1\) factors and a normal vector state \(\omega\) such that
\[
\beta_M(\omega;\mathcal M_A,\mathcal M_B,\mathcal M_C)=1.
\]

To see this, let \(N_i\subset\mathcal B(H_i)\), \(i=1,2,3\), be strongly stable type \(\mathrm{III}_1\) factors, for example hyperfinite type \(\mathrm{III}_1\) factors. On \(H=H_1\otimes H_2\otimes H_3\), define
\[
\mathcal M_A=N_1\otimes I_{H_2}\otimes I_{H_3},
\]
\[
\mathcal M_B=I_{H_1}\otimes N_2\otimes I_{H_3},
\qquad
\mathcal M_C=I_{H_1}\otimes I_{H_2}\otimes N_3.
\]
These algebras commute pairwise, and each is isomorphic to the corresponding factor \(N_i\).

Choose unit vectors \(\xi_i\in H_i\), and set
\(
\xi=\xi_1\otimes\xi_2\otimes\xi_3.
\)
Define the normal vector states
\(\omega(D)=\langle\xi,D\xi\rangle,\, D\in\mathcal B(H),
\)
and
\(
\omega_i(d)=\langle\xi_i,d\xi_i\rangle,\, d\in N_i,\, i=1,2,3.
\)
For self-adjoint contractions \(a\in N_1\), \(b\in N_2\), and \(c\in N_3\), we have
\(
\omega(a\otimes b\otimes c)
=\omega_1(a)\omega_2(b)\omega_3(c),
\)
where each factor belongs to \([-1,1]\). Thus, for any admissible local observables
\(
A_x=a_x\otimes I_{H_2}\otimes I_{H_3},\,
B_y=I_{H_1}\otimes b_y\otimes I_{H_3},\,
C_z=I_{H_1}\otimes I_{H_2}\otimes c_z,
\)
the expectation of the Mermin operator reduces to the classical scalar expression
\(
\omega(T)
=\frac12\bigl(
\alpha_0\beta_0\gamma_0
-\alpha_0\beta_1\gamma_1
-\alpha_1\beta_0\gamma_1
-\alpha_1\beta_1\gamma_0
\bigr),
\)
where
\(
\alpha_x=\omega_1(a_x),\,
\beta_y=\omega_2(b_y),\,
\gamma_z=\omega_3(c_z).
\)
The classical Mermin bound therefore gives \(|\omega(T)|\leq1\), and hence
\(
\beta_M(\omega;\mathcal M_A,\mathcal M_B,\mathcal M_C)\leq1.
\)
Conversely, choosing
\(
A_0=A_1=B_0=B_1=C_0=I_H,\, C_1=-I_H,
\)
gives \(T=I_H\). Therefore,
\(
\beta_M(\omega;\mathcal M_A,\mathcal M_B,\mathcal M_C)=1.
\)
\end{remark}

Strong stability of the individual local factors is therefore insufficient for maximal Mermin violation in every normal state. We next establish a strict upper bound under a factorization assumption on one pair of local algebras.

\begin{theorem}\label{thm:ACobstruction}
Let \(\mathcal M_A,\mathcal M_B,\mathcal M_C\subset\mathcal B(\mathcal H)\) be pairwise commuting von Neumann algebras with a common identity, and let \(\omega\) be a normal state on
\(\mathcal M_{ABC}
=\mathcal M_A\vee\mathcal M_B\vee\mathcal M_C.
\)
Assume that there exist two distinct parties \(J,K\in\{A,B,C\}\) such that
\(
\omega(xy)=\omega(x)\omega(y),\,
\text{for all }x\in\mathcal M_J,\ y\in\mathcal M_K.
\)
Then
\[
\beta_M(\omega;\mathcal M_A,\mathcal M_B,\mathcal M_C)
\leq\frac4{\sqrt5}<2.
\]
\end{theorem}
\begin{proof} By the symmetry among $\cM_A,\ \cM_B$ and $\cM_C$, it suffices to prove the case in which $\omega(ac)=\omega(a)\omega(c),\ \forall a\in\cM_A,c\in\cM_C$. 
 By Lemma~\ref{lem:symmetries}, it suffices to consider self-adjoint
unitaries for all six observables.

Hence assume $$A_i^{*}=A_i,\  A_i^2=I,\ B_j^{*}=B_j,\ B_j^2=I,\ C_k^{*}=C_k,\ C_k^2=I.$$

Set
\[
 L_A=\frac{[A_0,A_1]}{2i},\qquad
 L_B=\frac{[B_0,B_1]}{2i},\qquad
 L_C=\frac{[C_0,C_1]}{2i}.
\]
These are self-adjoint contractions and commute across different
parties,  i.e., $L_A^{*}=L_A,\ L_B^{*}=L_B,\ L_C^{*}=L_C$ and $[L_A,L_B]=[L_A,L_C]=[L_B,L_C]=0$.   Moreover, $\|[A_0,A_1]\|\le2\|A_0\|\|A_1\|=2,$  so $\|L_A\|\le1$. Expanding \eqref{eq:mermin3} and using
$A_i^2=B_j^2=C_k^2=I$ gives
\begin{equation}\label{eq:mer-square}
 T^2=I+L_AL_B+L_AL_C+L_BL_C.
\end{equation} And it is also easy to verify that $\{L_A,A_1\}=0,\ \{L_A,A_0\}=0$ by using the $A_i^2=I$ for $i\in\{0,1\}$. 
Moreover, $L_A$ commutes with every $B_j$ and $C_k$.
Each term of $T$ is a scalar multiple of some
$A_iB_jC_k$, and therefore
$TL_A+L_AT=0.$
Similarly, $TL_C+L_CT=0.$

We first note that, for any two anticommuting self-adjoint
contractions $P$ and $Q$ and any state $\omega$,  $\omega(P)^2+\omega(Q)^2\le 1$.
Indeed, for all $s,t\in\mathbb R$,
$(sP+tQ)^2
=s^2P^2+t^2Q^2
\le (s^2+t^2)I.$
The Cauchy--Schwarz inequality for states therefore yields
\[
|s\omega(P)+t\omega(Q)|^2
\le \omega\bigl((sP+tQ)^2\bigr)
\le s^2+t^2.
\]
Taking $s=\omega(P)$ and $t=\omega(Q)$, and writing
\[
r:=\omega(P)^2+\omega(Q)^2\ge 0,
\]
we obtain $r^2\le r$, which implies $r\le 1$.

Since $\|T\|\le 2$, both $T/2$ and $L_A$ are
self-adjoint contractions. Applying the preceding
inequality to $P=T/2$ and $Q=L_A$, and then to
$P=T/2$ and $Q=L_C$, gives
\[
\frac{\omega(T)^2}{4}+\omega(L_A)^2\le 1,
\qquad
\frac{\omega(T)^2}{4}+\omega(L_C)^2\le 1.
\]
Set
\[
m:=|\omega(T)|,\qquad
\alpha:=\omega(L_A),\qquad
\gamma:=\omega(L_C).
\]
Then
$\alpha^2,\gamma^2\le 1-\frac{m^2}{4},$
and consequently
$\alpha\gamma
\le |\alpha\gamma|
\le 1-\frac{m^2}{4}.$

The factorization assumption gives
$\omega(L_AL_C)=\omega(L_A)\omega(L_C)
=\alpha\gamma.$
Since $L_AL_B$ and $L_BL_C$ are self-adjoint
contractions, their expectations are at most $1$.
Using Eq. \eqref{eq:mer-square} and the
Cauchy--Schwarz inequality for states, we obtain
\[
\begin{aligned}
m^2
&=|\omega(T)|^2\\
&\le \omega(T^2)\\
&=1+\omega(L_AL_B)+\omega(L_AL_C)+\omega(L_BL_C)\\
&\le 3+\alpha\gamma\\
&\le 4-\frac{m^2}{4}.
\end{aligned}
\]
Thus $5m^2\le 16$, and hence
$|\omega(T)|\le \frac{4}{\sqrt{5}}.$
Taking the supremum over all local self-adjoint
unitaries and applying Lemma \ref{lem:symmetries} yields
\[
\beta_M(\omega;\mathcal M_A,\mathcal M_B,\mathcal M_C)
\le \frac{4}{\sqrt{5}}<2.
\]\end{proof}

\begin{example} There exist three strongly stable type $\III_1$ factors satisfying
\begin{equation}\label{eq:triduality}
 \cM_A'=\cM_B\vee\cM_C,\qquad
 \cM_B'=\cM_A\vee\cM_C,\qquad
 \cM_C'=\cM_A\vee\cM_B,
\end{equation}
for which maximal Mermin violation does not hold in every normal
state. 

Indeed, in the Hilbert space $\cH=\cH_L\otimes \cH_R$, let $\cM_A=\cM_L\vN I_R$, $\cM_B=\cM_L'\vN \cM_R$ and $\cM_C=I_L\otimes \cM_R'$, where $\cM_L$ and $\cM_R$ act on $\cH_L$ and $\cH_R$ respectively.
Take $\cM_L,\cM_R$ to be hyperfinite type $\III_1$
factors in standard form. Then $\cM_A,\cM_B,\cM_C$ are strongly stable
type $\III_1$ factors. Since
$\cM_L\vee \cM_L'=\mathcal B(\cH_L)$, and similarly on the right,
direct calculation gives
\begin{align*}
 \cM_B\vee\cM_C&=\cM_L'\vN\mathcal B(\cH_R)=\cM_A',\\
 \cM_A\vee\cM_C&=\cM_L\vN \cM_R'=\cM_B',\\
 \cM_A\vee\cM_B&=\mathcal B(\cH_L)\vN \cM_R=\cM_C'.
\end{align*}
Every normal product state $\omega=\omega_L\otimes\omega_R$ satisfies $\omega(ac)=\omega(a)\omega(c)$, so Theorem~\ref{thm:ACobstruction} excludes $\mer=2$
in these states. 
\end{example}

Finally, the most important question is on what $\III_1$ factors Mermin inequalities can be maximally violated for all normal states. We now give a sufficient condition for maximal Mermin violation in every normal state.

Let
\(
\mathfrak h=(\mathbb C^2)^{\otimes3},\,
|GHZ\rangle
=\frac{|000\rangle+|111\rangle}{\sqrt2},
\)
and define the incomplete infinite tensor product
\(
\mathcal H_G
=\bigotimes_{m=1}^{\infty}
(\mathfrak h,|GHZ\rangle),
\,
\Omega_G
=\bigotimes_{m=1}^{\infty}|GHZ\rangle.
\)
Here \(|GHZ\rangle\) denotes the single-coordinate GHZ vector, whereas \(\Omega_G\) denotes the reference vector in \(\mathcal H_G\). For each \(m\), let
\(
\iota_m:\mathcal B(\mathfrak h)\longrightarrow\mathcal B(\mathcal H_G)
\)
denote the embedding acting on the \(m\)-th coordinate and as the identity on all other coordinates. Define
\(
\alpha_A(a)=a\otimes I_2\otimes I_2,\,
\alpha_B(a)=I_2\otimes a\otimes I_2,\quad
\alpha_C(a)=I_2\otimes I_2\otimes a.
\) for \(a\in M_2(\mathbb C)\).
For \(J\in\{A,B,C\}\), set
\(
\mathcal R_J^G
=
\left(
\bigcup_{m\geq1}
\iota_m\bigl(\alpha_J(M_2(\mathbb C))\bigr)
\right)''.
\)
These three von Neumann algebras commute pairwise.

\begin{definition}\label{def:GHZembed}
Let $\cM_A, \cM_B,\cM_C\subset B(\cH)$ be pairwise commuting von Neumann algebras. We say that the triple $(\cM_A, \cM_B,\cM_C)$ admits a joint GHZ tail embedding  
if there exist a Hilbert space $\cK$ and a unitary
$W:\cH\to\cK\otimes \cH_G$ such that
\begin{equation}\label{eq:GHZembedding}
 \id\otimes\cR_A^G\subset W\cM_A W^*,\qquad
 \id\otimes\cR_B^G\subset W\cM_B W^*,\qquad
 \id\otimes\cR_C^G\subset W\cM_C W^*.
\end{equation}
\end{definition}

\begin{theorem}\label{thm:GHZnormal}
If mutually commuting von Neumann algebras $\cM_
A, \cM_B, \cM_C$ satisfy the joint GHZ tail embedding condition, then there is a
common sequence $(T_m)$ of Mermin operators such that 
$T_m\sot2 I$. Consequently, every global normal state $\omega$ on $\cM_{\rm ABC}$
satisfies
\[
 \mer(\omega;\cM_A,\cM_B,\cM_C)=2.
\]
\end{theorem}

\begin{proof}Let \(X\) and \(Y\) be the Pauli matrices, and define
\[
t=\frac12\bigl(
X\otimes X\otimes X
-X\otimes Y\otimes Y
-Y\otimes X\otimes Y
-Y\otimes Y\otimes X
\bigr)
\]
on \(\mathfrak h=(\mathbb C^2)^{\otimes3}\). Then
\(
t|\mathrm{GHZ}\rangle=2|\mathrm{GHZ}\rangle,
\qquad \|t\|\leq2.
\)

Using the coordinate embeddings defined above, set
\[
\begin{aligned}
A_{0,m}^G&=\iota_m(\alpha_A(X)),
& A_{1,m}^G&=\iota_m(\alpha_A(Y)),\\
B_{0,m}^G&=\iota_m(\alpha_B(X)),
& B_{1,m}^G&=\iota_m(\alpha_B(Y)),\\
C_{0,m}^G&=\iota_m(\alpha_C(X)),
& C_{1,m}^G&=\iota_m(\alpha_C(Y)).
\end{aligned}
\]
These are local self-adjoint unitaries. Their associated Mermin operator is
\(
T_m^G=\iota_m(t),
\)
and therefore \(T_m^G\Omega_G=2\Omega_G\).

Let \(\mathcal D\) be the linear span of the elementary tensors
\(
\zeta=\bigotimes_{k=1}^{\infty}\zeta_k
\)
such that \(\zeta_k=|\mathrm{GHZ}\rangle\) at all but finitely many coordinates. By the definition of the incomplete infinite tensor product, \(\mathcal D\) is dense in \(\mathcal H_G\).

For such an elementary tensor, let
\(
F=\{k:\zeta_k\neq|\mathrm{GHZ}\rangle\},
\, N=\max(F\cup\{0\}).
\)
If \(m>N\), then \(T_m^G\) acts on a coordinate whose vector is \(|\mathrm{GHZ}\rangle\), so
\(
T_m^G\zeta=2\zeta.
\)
By linearity, for every \(\zeta\in\mathcal D\) there exists \(N(\zeta)\) such that
\(
T_m^G\zeta=2\zeta\,\text{whenever }m>N(\zeta).
\)
Since \(\sup_m\|T_m^G\|\leq2\), convergence on \(\mathcal D\) extends by density to all of \(\mathcal H_G\). Hence
\(
T_m^G\xrightarrow{\mathrm{s}}2I_{\mathcal H_G}.
\)

By the joint GHZ tail embedding assumption, there exists a unitary
\(
W:\mathcal H\longrightarrow\mathcal K\otimes\mathcal H_G
\)
satisfying the inclusions in Definition \ref{def:GHZembed}. Define
\[
\begin{aligned}
A_{x,m}&=W^*(I_{\mathcal K}\otimes A_{x,m}^G)W\in\mathcal M_A,\\
B_{y,m}&=W^*(I_{\mathcal K}\otimes B_{y,m}^G)W\in\mathcal M_B,\\
C_{z,m}&=W^*(I_{\mathcal K}\otimes C_{z,m}^G)W\in\mathcal M_C.
\end{aligned}
\]
The associated Mermin operator is
\(
T_m=W^*(I_{\mathcal K}\otimes T_m^G)W.
\)
Convergence on elementary tensors, together with uniform boundedness, gives
\(
I_{\mathcal K}\otimes T_m^G
\xrightarrow{\mathrm{s}}
2I_{\mathcal K\otimes\mathcal H_G}.
\)
Consequently,
\(
T_m\xrightarrow{\mathrm{s}}2I_{\mathcal H}.
\)
The conclusion now follows from Theorem \ref{thm:Mequiv}.
\end{proof}

The following construction gives examples with type \texorpdfstring{$\III_1$}{III1} factors satisfying the conditions in the above theorem. Therefore, we show a class of $\III_1$ factors on which Mermin inequalities are maximally violated for all normal states.

\begin{proposition}\label{prop:GHZfactors}Let \(\mathcal R_A^G,\mathcal R_B^G,\mathcal R_C^G\subset\mathcal B(\mathcal H_G)\) be the von Neumann algebras defined above. Each is isomorphic to the hyperfinite type \(\mathrm{II}_1\) factor \(\mathcal R\). 
Let $\cN_A,\cN_B,\cN_C$ be type $\III_1$ factors acting on
$\cK_A,\cK_B,\cK_C$, respectively. On
$\cK_A\otimes \cK_B\otimes \cK_C\otimes \cH_G$, set
\begin{align*}
 \cM_A&=\cN_A\vN\mathbb C\id\vN\mathbb C\id\vN\cR_\cA^G,\\
  \cM_B&=\mathbb C\id\vN \cN_B\vN\mathbb C\id\vN\cR_B^G,\\
  \cM_C&=\mathbb C\id\vN\mathbb C\id\vN \cN_C\vN\cR_C^G.
\end{align*}
Then these are three strongly stable type $\III_1$ factors, and every
global normal state satisfies $\mer=2$.
\end{proposition}

For each party, the restriction of the reference vector state to every finite-coordinate local matrix algebra is the normalized trace. The corresponding representation on \(\mathcal H_G\) is an amplification of the tracial infinite tensor product representation of \(M_2(\mathbb C)\). Thus each \(\mathcal R_J^G\) is a hyperfinite type \(\mathrm{II}_1\) factor. 

{\it Application to quantum field theory and comparison of the results} Let $O\mapsto\mathfrak A(O)$ be a net satisfying locality. For
pairwise spacelike separated regions $O_A,O_B,O_C$, one may take
$\cM_A=\mathfrak A(O_A)$, $\cM_B=\mathfrak A(O_B)$, and
$\cM_C=\mathfrak A(O_C)$. Applying the results above requires
verification of one of the following additional structures:

 For Mermin correlations, the local triple must satisfy the joint
embedding condition of Definition~\ref{def:GHZembed}, or one must
directly construct an asymptotic GHZ stabilizer sequence as in
Theorem~\ref{thm:Mequiv}.

The original two-party argument for complementary wedges does not
verify these conditions for arbitrary triples of regions. For strictly
separated regions satisfying an appropriate split property, the
existence of normal product states instead obstructs maximal violation
in every state. Thus the type $\III_1$ property of local algebras alone
does not establish general tripartite maximality in QFT.

\section{N-partite case}
\label{sec:npartite}

We extend the structural results of the tripartite case to every finite
number \(n\ge3\) of pairwise commuting von Neumann algebras.
Here the proof of all conclusions are omitted since they are similar to those of the tripartite case.

We first build the multipartite Mermin inequality for the commuting algebra model. Fix an integer \(n\geq3\). Let
\[
 \cM_{A_1},\ldots,\cM_{A_n}\subset B(\cH),\qquad
 [\cM_{A_j},\cM_{A_\ell}]=0\quad(j\neq\ell),
\]
be von Neumann algebras with the same identity \(\id\), acting on a
common separable Hilbert space \(\cH\).  For \(x\in\{0,1\}\), the observable \(A_{j,x}\in\cM_{A_j}\)
is a self-adjoint contraction. Write \(A=(A_{j,x})_{j,x}\), and set
\begin{equation}\label{eq:constants}
 c_n=2^{\lfloor n/2\rfloor},\qquad N_n=2^{n-1},\qquad q_n=\frac{N_n}{c_n}.
\end{equation}
We use the real-part convention and define
\begin{equation}\label{eq:Tn}
 T_n(A)=\frac{1}{c_n}\Rea\prod_{j=1}^{n}(A_{j,0}+iA_{j,1}),
 \qquad \Rea D=\frac{D+D^*}{2}.
\end{equation}
For a state \(\omega\) on
\(\cM_{A_1}\vee\cdots\vee\cM_{A_n}\), put
\begin{equation}\label{eq:beta}
\mer(\omega;\cM_{A_1},\ldots,\cM_{A_n})=\sup_A|\omega(T_n(A))|.
\end{equation}
Let
\[
 \Even=\{S\subseteq\{1,\ldots,n\}:|S|\text{ is even}\},\qquad
 R_S(A)=(-1)^{|S|/2}\prod_{j\in S}A_{j,1}\prod_{j\notin S}A_{j,0}.
\]
Commutation between different parties gives
\begin{equation}\label{eq:expansion}
 T_n(A)=\frac{1}{c_n}\sum_{S\in\Even}R_S(A),\qquad |\Even|=N_n.
\end{equation}

\begin{theorem}\label{thm:main}
Suppose that at least \(k\) of the \(n\) local algebras \(\cM_{A_j}\)
are abelian, where \(0\leq k\leq n\). Then, for every admissible \(A\),
\begin{equation}\label{eq:norm-bound}
 \|T_n(A)\|\leq b_{n,k},
\end{equation}
where
\begin{equation}\label{eq:b}
 b_{n,k}=\max\{1,2^{-k/2}q_n\}
       =\max\{1,2^{\,n-1-\lfloor n/2\rfloor-k/2}\}.
\end{equation}
Consequently, for every state \(\omega\),
\begin{equation}\label{eq:interval}
 1\leq\mer(\omega;\cM_{A_1},\ldots,\cM_{A_n})\leq b_{n,k}.
\end{equation}
Writing \(r=n-k\), the same constant is
\begin{equation}\label{eq:parity}
 b_{n,k}=
 \begin{cases}
 2^{\max\{r-1,0\}/2},&n\text{ odd},\\
 2^{\max\{r-2,0\}/2},&n\text{ even}.
 \end{cases}
\end{equation}
\end{theorem}

\begin{lemma}
\label{lem:symmetries1}
Let  $\cM_{A_1},\cdots,\cM_{A_n}\subset B(\cH)$ be pairwise commuting von Neumann algebras with a common identity. Denote for each $j\in\{1,\cdots,n\}:$
$M_j=\{A\in \cM_{A_j}|A^{*}=A\},\ K_j=\{A\in M_j|\|A\|\le1\}.$ For a normal state $\omega$ on $
\mathcal M=\bigvee_{j=1}^{n}\mathcal M_{A_j}$ and the $n$-partite Mermin operator 
$$T_n({\bf A})=\frac{1}{c_n}\sum_{S\in Even}(-1)^{|S|/2}\prod_{j\in S}A_{j,1}\prod_{j\notin S}A_{j,0},$$
one has 
$$\sup_{{\bf A}\in \prod (K_j\times K_j)}|\omega(T_n({\bf A}))|=\sup_{{\bf A}\in\prod (\text{ext} K_j\times\text{ext}K_j)}|\omega(T_n({\bf A}))|.$$
Moreover, for each $j$, $\text{ext}K_j=\{A\in K_j:A^2=I\},$ i.e., the extreme points of $K_j$ are precisely the self-adjoint unitaries in $\cM_{A_j}$. Consequently, in computing $\beta_M(\omega)$, all $2n$ self-adjoint contractions may be restricted to self-adjoint unitaries.
\end{lemma}

\begin{proposition}\label{prop:bounds}
Let $\cM_{A_1},\cdots,\cM_{A_n}\subset B(\cH)$ be pairwise commuting von Neumann algebras, and set
\(\mathcal M=\bigvee_{j=1}^{n}\mathcal M_{A_j}.\)
Let \(\omega\) be a normal state on \(\mathcal M\). For the normalized $n$-party Mermin operator $T_n({\bf A})$, the maximal expectation value over all self-adjoint contractions is universally bounded by
$$\beta_M(\omega)=\sup_{\bf A}|\omega(T_n({\bf A}))|\le q_n,$$
where the quantum bound $q_n$ is given by 
\[
  q_n
 =\begin{cases}
  2^{(n-1)/2},&n\ \text{odd},\\
  2^{(n-2)/2},&n\ \text{even}.
 \end{cases}
\]
The universal quantum bound is sharp.
\end{proposition}

We first show how a common sequence characterizes maximality in all normal states.
\begin{definition}
For each $m\in\mathbb N$, let
$A^{(m)}
=
\bigl(A_{j,x}^{(m)}\bigr)_{
1\le j\le n,\;x\in\{0,1\}},$
where each $A_{j,x}^{(m)}\in\cM_{A_j}$ is a
self-adjoint unitary.
We say that the sequence $(A^{(m)})_{m\in\mathbb N}$
satisfies the \emph{asymptotic $n$-party GHZ relations} if
\begin{equation}\label{eq:asymptotic}
R_S^{(m)}
:=
R_S(A^{(m)})
\sot I
\qquad
(S\in\mathrm{Even}).
\end{equation}
Here $n$ is fixed, and the convergence is taken
as $m\to\infty$.
\end{definition}

\begin{theorem}\label{thm:common} Let $\cM_{A_1},\cdots,\cM_{A_n}\subset B(\cH)$ be pairwise commuting von Neumann algebras, and set 
\(
\mathcal M=\bigvee_{j=1}^{n}\mathcal M_{A_j}.
\) For the normalized  $n$-party Mermin functional defined in Eq.\,(\ref{eq:beta}), the following conditions are equivalent:
\begin{enumerate}[label=\textup{(\roman*)}]
 \item \(\mer(\omega)=q_n\) for every normal state \(\omega\) on \(\mathcal M\).
 \item \(\mer(\rho)=q_n\) for at least one faithful normal state \(\rho\) on \(\mathcal M\).
 \item There is a sequence of local symmetries with
       \(T_n(A^{(m)})\sot q_n\id\).
 \item There is a sequence of local symmetries satisfying
       \eqref{eq:asymptotic}.
\end{enumerate}
Replacing strong convergence by ultraweak convergence in
\textup{(iii)} gives an equivalent condition.
\end{theorem}

\begin{theorem}\label{thm:necessary}
Let $\cM_{A_1},\cdots,\cM_{A_n}\subset B(\cH)$ be pairwise commuting von Neumann algebras. Suppose that each $\cM_{A_j}$ is a factor and that
the $n$-tuple $(\cM_{A_1},\ldots,\cM_{A_n})$
satisfies the equivalent conditions of
Theorem~\ref{thm:common}.
Then each local factor $\cM_{A_j}$ is strongly stable; that is, 
\[\cM_{A_j}\cong\cM_{A_j}\vN\cR\qquad(1\le j\le n),
\] where $\cR$ is the hyperfinite type $II_1$ factor.
\end{theorem}

Similar to the tripartite case, we can obtain a sufficient GHZ tail embedding condition for all normal state maximal violation. 
Let the single-coordinate $n$-party GHZ state be denoted by 
$|GHZ_n\>=\frac{1}{\sqrt{2}}(|0\>^{\otimes n}+|1\>^{\otimes n})\in (\C^2)^{\otimes n}.$
Consider the incomplete infinite tensor product Hilbert space
\begin{equation}\label{eq:GHZ-space}
 \cH_{G,n}=\bigotimes_{m=1}^{\infty}
     \bigl((\mathbb C^2)^{\otimes n},|GHZ_n\>\bigr),
 \qquad \Omega_{G,n}=\bigotimes_{m=1}^{\infty}|GHZ_n\>
\end{equation}
Here \(\Omega_{G,n}\) is the reference vector in \(\mathcal H_{G,n}\). Let
\(
\iota_m^{(n)}:
\mathcal B((\mathbb C^2)^{\otimes n})
\longrightarrow\mathcal B(\mathcal H_{G,n})
\)
denote the \(m\)-th coordinate embedding. For \(1\leq j\leq n\), define
\(
\alpha_j^{(n)}(a)
=
I_2^{\otimes(j-1)}
\otimes a\otimes
I_2^{\otimes(n-j)},\, a\in M_2(\mathbb C),
\)
and set
\[
\mathcal R_j^{G,n}
=
\left(
\bigcup_{m\geq1}
\iota_m^{(n)}
\bigl(\alpha_j^{(n)}(M_2(\mathbb C))\bigr)
\right)''.
\]
These \(n\) von Neumann algebras commute pairwise.

\begin{definition}\label{def:embedding}
A family of von Neumann algebras $(\cM_{A_1},\cdots,\cM_{A_n})$ is said to admit a joint GHZ tail embedding if there exist 
a Hilbert space \(\cK\) and a unitary
\(W:\cH\to\cK\otimes\cH_{G,n}\) such that
\begin{equation}\label{eq:embedding}
 \id_{\cK}\otimes\cR_j^{G,n}\subseteq W\cM_{A_j}W^*
 \qquad(1\le j\le n).
\end{equation}
\end{definition}

\begin{theorem}\label{thm:embedding}
If the configuration of von Neumann algebras $(\cM_{A_1},\cdots,\cM_{A_n})$ admits a joint GHZ tail embedding condition as in (\ref{eq:embedding}), then it satisfies the equivalent conditions of
Theorem~\ref{thm:common}. In particular,
\(\mer(\omega;\cM_{A_1},\ldots,\cM_{A_n})=q_n\) for every normal state $\omega$ on
$\mathcal M_{A_1}\vee\cdots\vee\mathcal M_{A_n}$.
\end{theorem}

\begin{proposition}\label{prop:GHZ-factors}
Let \(\mathcal R_1^{G,n},\ldots,\mathcal R_n^{G,n}\subseteq\mathcal B(\mathcal H_{G,n})\) be the von Neumann algebras defined above.  Each $\cR^{G,n}_j$ is isomorphic to the hyperfinite type $\II_1$ factor $\cR$. 
Let \(N_j\subseteq\mathcal B(\cK_j)\) be type \(\III_1\) factors acting on $\cK_j$. On the composite Hilbert space
 \((\bigotimes_{j=1}^n\cK_j)\otimes\cH_{G,n}\), 
 set
\[
 \cM_{A_j}=\id\otimes\cdots\otimes N_j\otimes\cdots\otimes\id
                \vN\cR_j^{G,n}
\]
Then the algebras
$\mathcal M_{A_1},\ldots,\mathcal M_{A_n}$
are pairwise commuting, strongly stable
type $\mathrm{III}_1$ factors, and
\(
\beta_M(\omega;\mathcal M_{A_1},\ldots,\mathcal M_{A_n})
=q_n
\)
for every normal state $\omega$ on
$\mathcal M_{A_1}\vee\cdots\vee\mathcal M_{A_n}$.
\end{proposition}

\section{Discussion}
\label{sec:interpretation}

\begin{openproblem}[A state-independent criterion under standard form
and multiparty duality]\label{op:duality}
In the bipartite case the criterion of Summers and Werner is an exact
characterization: for a factor $\cM$ with separable predual acting on a
Hilbert space on which it has a cyclic and separating vector,
\begin{equation}\label{eq:SWcriterion}
 \beta_2(\omega;\cM,\cM')=\sqrt2
 \quad\text{for every normal state }\omega
 \iff \cM\cong\cM\vN\cR .
\end{equation}
The bipartite criterion uses a standard representation and the pairing of an algebra with its full 
commutant. In QFT, Haag duality relates this commutant to the algebra of the causal complement. 
For three algebras, Theorems~\ref{thm:necessary3}
and~\ref{thm:GHZnormal} give
\[
 \text{individual strong stability}
 \ \Longleftarrow\ \text{maximality}
 \ \Longleftarrow\ \text{joint GHZ tail embedding}.
\]
Individual strong stability does not imply maximality. The joint GHZ tail embedding condition is sufficient; its necessity is not established here.  The problem is to close the gap under some suitable hypotheses that correspond to the two features above.
For instance, in the triparty case, we want to know:
\emph{Assume that $\cM_\cA,\cM_\cB,\cM_\cC$ are pairwise commuting
factors admitting a common cyclic and separating vector $\Omega$, and
that the triparty duality relations hold. Is
$\mer(\omega)=2$ for all normal states characterized by an intrinsic
algebraic property of the triple?}
 \end{openproblem}

\begin{openproblem}[Other multipartite inequalities: MABK and
Svetlichny]\label{op:inequalities}
The functional \eqref{eq:Tn} is a real-part Mermin polynomial, and
with $c_n=2^{\lfloor n/2\rfloor}$ its maximal quantum value is
$q_n=2^{n-1-\lfloor n/2\rfloor}$. The standard MABK
normalization~\cite{Mermin,Ardehali,BK} has maximal quantum value
$2^{(n-1)/2}$ for every $n$. The two numbers coincide when $n$ is odd,
whereas
\begin{equation}\label{eq:evengap}
 q_n=2^{n/2-1}<2^{(n-1)/2}\qquad(n\ \text{even}).
\end{equation}
With the local bound normalized to \(1\), the real-part Mermin functional considered here has maximal quantum value
\(
q_n=2^{n-1-\lfloor n/2\rfloor},
\)
whereas the standard MABK family has maximal quantum value \(2^{(n-1)/2}\). For even \(n\),
\(
q_n=\frac1{\sqrt2}\,2^{(n-1)/2}.
\)
Thus the real-part functional has a smaller maximal quantum-to-classical ratio. This numerical comparison does not by itself establish an ordering of the two families by their ability to detect nonlocality.

For even $n$, the standard MABK polynomial involves a relative phase. Any extension must 
specify whether this phase is absorbed into the GHZ state or into the local observables. Thus the asymptotic relations
\eqref{eq:asymptotic} have no immediate MABK analogue.

\begin{enumerate}[label=(\alph*)]
 \item \emph{An MABK counterpart of Theorem~\ref{thm:common}.} Does
 the equivalence of Theorem~\ref{thm:common} admit an MABK version for
 even $n$, with \eqref{eq:asymptotic} replaced by phase-rotated
 asymptotic stabilizer relations built from the optimal settings?
 \item \emph{Maximizing configurations.} Which configurations of commuting local algebras yield maximal MABK violation in every normal state? Is a joint GHZ tail embedding necessary for this property?
\end{enumerate}

A second family is Svetlichny's~\cite{Svetlichny}, designed to certify
genuine tripartite nonlocality: its bound is $4$ for every biseparable
model, its maximal quantum violation is $4\sqrt2$, attained by the GHZ
state, and the $n$-party versions are~\cite{SS}. In the commuting-algebra model
the algebraic counterpart of biseparability is the closed convex hull
of the normal states factorizing across one of the three bipartitions
$A|BC$, $B|AC$, $C|AB$. Whether the bound $4$ holds for this class is
not automatic, because the standard proof exploits the local hidden
variable structure rather than commutation of the algebras.

The same questions can be asked for the full correlation hierarchy in
the commuting-operator model, for the Bell-type inequalities of the
companion paper~\cite{YHH}, and for the algebraic analogues of the
device-independent protocols whose security rests on these
inequalities~\cite{BCPSW}.
\end{openproblem}

\section*{Acknowledgements}

 The work was supported by the National Natural Science Foundation of
China (Grant Nos.~12271394, 12601237). This work was funded by the China
Scholarship Council.

\end{document}